# Control coordination between DFIG-based wind turbines and synchronous generators for optimal primary frequency response

Samaneh Morovati, Héctor Pulgar-Painemal
*Department of Electrical Engineering and Computer Science*
*The University of Tennessee, Knoxville*
Knoxville, TN, USA
smorovat@utk.edu, hpulgar@utk.edu

*Abstract*—This paper proposes a novel coordinating mechanism between synchronous generators (SGs) and wind turbines (WTs) based on doubly-fed induction generators (DFIGs) for enhanced primary frequency regulation. WTs are urged to participate on frequency regulation, specially if wind power penetration keeps increasing. WTs control support is possible, but it is transient due to the WTs lack of energy storage. This drawback can result in either a further delayed response from the governors of SGs or further frequency decay when WTs support is over. The proposed coordination attempt to tackle this issue. An artificial neural network (ANN) is used to obtain an optimal coordination signal to improve frequency response. As a proof of concept, the proposed coordination is tested on a 9-bus test system that includes a wind farm with 5 WTs. Simulation results show that frequency nadir is reduced in about 22% and rates of change of the system frequency (RoCoF) in about 29.5%. Further work is needed to validate this concept in large-scale systems, but the development and results obtained so far are promising to strengthen power systems.

## Nomenclature

**Physical Variables**

| | |
|---|---|
| $\delta$ | Loadability angle of a SG |
| $\omega$ | Electrical angular speed of a SG |
| $\omega_s$ | Synchronous angular speed of a SG |
| $E_{fd}$ | Field voltage of a SG |
| $E_q'$ | q-axis transient internal voltage of a SG |
| $H$ | Inertia constant of a SG |
| $I_q, I_d$ | q-axis and d-axis stator currents of a SG |
| $K_A$ | Regulator gain of the IEEE Type-1 exciter of a SG |
| $K_E$ | Gain of the IEEE Type-1 exciter of a SG |
| $K_F$ | Feedback gain of the IEEE Type-1 exciter of a SG |
| $P_{ref}$ | Active power reference of a WTG's speed controller |
| $Q_{ref}$ | Reference of the reactive power controller of a SG or DFIG |
| $T_A$ | Regulator time constant of the IEEE Type-1 exciter of a SG |
| $T_E$ | Time constant of the IEEE Type-1 exciter of a SG |
| $T_F$ | Feedback time constant of the IEEE Type-1 exciter of a SG |
| $T_m$ | Mechanical torque (SG or WTG) |
| $T_M$ | Mechanical torque of a WTG |
| $T_d'$ | d-axis transient time constant of a SG |
| $V_R$ | Voltage regulator output of a SG |
| $X_q, X_d$ | q-axis and d-axis steady-state reactance of a SG |
| $X_q', X_d'$ | Transient q-axis and d-axis reactance of a SG |

This work made use of Engineering Research Center shared facilities supported by the Engineering Research Center Program of the National Science Foundation and the Department of Energy under NSF Award No. EEC-1041877 and the CURENT Industry Partnership Program.

## I. Introduction

In recent years, there is a widespread usage of renewable resources as an alternative for powering the grid. These resources are mostly connected to the grid with power electronic interfaces and are completely decoupled from the grid. Among the growing options of asynchronous converter-interfaced generation, wind power technology continues to rise [1]. DFIG-based WTs are the mainstream for wind generation, due to their competitive cost and fast response during transient events such as frequency excursions [2], [3], [4]. A power imbalance event, caused by a generator trip or a load change in the grid, can cause unacceptably large excursions and RoCoF [5]. This phenomena calls the need to prioritize primary frequency control response [6], [7]. To overcome this problem, these renewable resources require to contribute to frequency control alongside the conventional synchronous plants with fast ramp rates and natural response of the frequency-dependant loads [2]. There are several papers that have proposed various control structures for wind farm contribution to frequency control.

Various control mechanisms and techniques for primary frequency control of WTs in an interconnected power system have been proposed [8]. As a primary frequency control methods that use WTs to contribute on grid frequency control, pitch control and realising kinetic energy of wind turbine based control are those that are common [9]. Beside these primary methods, in order to relieve the shortage of inertia and frequency excursion, virtual inertia frequency control strategy based on a frequency error integral loop was proposed in [10]. Since the frequency excursion can be caused by a generator trip or a load change in the grid that will act as large disturbance and will affect system reliability [11], a novel idea using model reference control method in order to capture the desired inertia response during a large disturbance in the microgrids was expressed in [12]. However, obtaining an optimal control signal is another concern among the proposed control strategies. Many investigations have suggested ways to increase the participation of wind farm and have proposed optimal control strategies for the system frequency control [13], [14], [15]. Providing an optimal control is one of the key frameworks for the frequency regulation problem proposed in [16]. Additionally, optimal decentralized primary frequency control in which every generator and load makes its decision based on local frequency sensing could be recommended [17]. An optimal control method for variable speed wind turbine can use as a temporary frequency control based on minimizing cost of energy [18]. However, there is still a concern that the power imbalance might hit back when the WTs frequency control support is ended [19]. This will require changes in the way the grid frequency is controlled and provide a permanent frequency support for the grid.

The change in output power at SGs resulting from the frequency excursion is controlled by their governors [20]. This primary frequency control, also known as droop control, have some time delay action in which a speed governor tune the generation power based on local frequency feedback [17]. Then the secondary control loop set back the frequency to the desired value by adjusting valves or power set-points [21]. To provide a solution for frequency support and tackle governor delays, the idea of a coordinating mechanism by preserving power of DFIGs and governor actions, which is based on grid frequency itself, is highly significant. In this case, a variable coefficient combined with virtual inertia and primary frequency control method was provided for DFIGs coordinating with diesel generators to participate in microgrid frequency regulation in [22]. In this paper, we provide a novel idea of a coordinating mechanism between DFIG-based WTs and SGs to enhance primary frequency regulation. The proposed optimal control is obtained through an ANN. As an advantage of this approach is its potential to be extended for complex large scale system, such as actual power grids, to obtain coordinated control signal for optimal primary frequency response.

This paper is organized as follows. Section II briefly introduces the primary objective for frequency response and includes model description. Section III provides the novel idea of primary frequency control and specified target for controller design procedure. Section IV provides the novel idea of designing neural network controller based on DFIG based WTs and synchronous generator governors. Comprehensive simulation results are presented in this section, followed by the conclusion in Section V.

## II. PRELIMINARIES AND MODEL DESCRIPTION

In this section we describe the required preliminaries of the model and the main objective of our work, which is primary frequency control. This type of control refers to the system frequency stabilization right after a power imbalance caused by, for example, a generator outage. The key components during this transient behavior are the governors of SGs, which can act very slowly specially when the generation is highly based on thermal power plants. The slow response of governors together with an initial small frequency deviation result in an almost linear frequency decay right after a disturbance, e.g., generator outage. This part is known as inertial response as kinetic energy is being drained from the generator masses to transiently balance the power consumption (see Fig. 1). After the inertial response, with a larger frequency deviation, the governors become more active and are able to stabilize the system frequency by adjusting the power output of the generators. Due to both droop characteristic in governors and frequency dependency in loads, generation and consumption is balanced and the system frequency is stabilized at a frequency different from its initial value. This part is known as governor response. Ideally, after the disturbance, the frequency evolves from its initial to its final value in a critically damped fashion, without overshooting. This overshooting is undesired as it can create an excessive frequency deviation (frequency nadir) that can trigger load shedding mechanisms. The aforementioned control structures in WTs aim to reduce this frequency nadir as this increases electric service continuity and system reliability. This paper is proposing a new concept for the frequency excursion control by considering a coordinating mechanism between governors of SGs and DFIG based WTs. As an advantage of this approach is its potential to extend for complex large scale systems, such as power grids to provide the proper control signal for the primary frequency control.

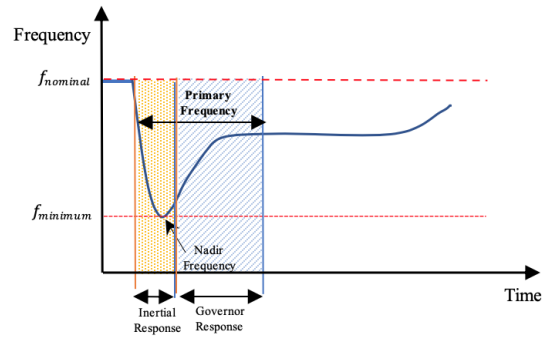

Fig. 1. Time frame frequency response.

The DFIG-based WTs are represented by a two-axis model that retains the dynamics of the rotor flux linkages and assumes that stator flux linkages can change instantaneously. To extract maximum power from an incoming wind going through the turbine blades, active power in WTs is controlled to follow the maximum power point tracking (MPPT) curve. In the case of reactive power, this is controlled to follow a reference given by the supervisory voltage controller of the wind farm.

Both active and reactive power are controlled through the adjustment of the DFIG rotor voltage and use PI controllers with a fast current loop and a slow power loop. More details about the model, references and technical limits are presented in reference [23]. The participation scheme for frequency regulation through the inertial response is illustrated in Fig. 2. This participation basically uses the kinetic energy stored at a wind turbine mass to inject more or less power in case of under or over frequency, respectively [24]. A washout filter is used to react only during the first moment of the deviation allowing a steady-state frequency different than the rated one [25].

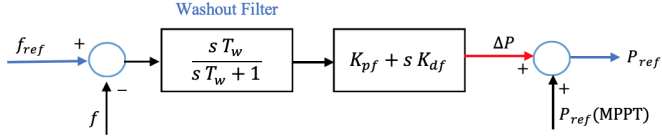

Fig. 2. WT control scheme for frequency support through inertial response.

The SGs are described by a one-axis model together with an IEEE Type-1 exciter for voltage control [26], and an IEESGO governor for frequency regulation [27]. The one-axis model for the SGs with IEEE Type-1 exciter is described by the following set of differential-algebraic equations (DAEs) [28]:

$$T'_{do}\frac{dE'_q}{dt} = -E'_q - (X_d - X'_d)I_d + E_{fd} \quad (1)$$

$$\frac{d\delta}{dt} = \omega - \omega_s$$

$$\frac{2H}{\omega_s}\frac{d\omega}{dt} = T_M - E'_q I_q - (X_q - X'_d)I_d I_q - T_{FW} \quad (2)$$

$$T_E\frac{dE_{fd}}{dt} = -(K_E + S_E(E_{fd}))E_{fd} + V_R \quad (3)$$

$$T_A\frac{dV_R}{dt} = -V_R + K_A\left(R_f - \frac{K_F}{T_F}E_{fd} + (V_{ref} - V_t)\right) \quad (4)$$

$$T_F\frac{dR_f}{dt} = -R_f + \frac{K_F}{T_F}E_{fd} \quad (5)$$

The governor is represented by the following set of DAEs:

$$T_1\frac{y_1}{dt} = -y_1 + K_1(\omega - \omega_s + u_c)/\omega_s \quad (6)$$

$$T_3\frac{y_3}{dt} = -y_3 + y_1 \quad (7)$$

$$T_4\frac{T_M}{dt} = -T_M + P_C - y_2 \quad (8)$$

$$y_{2i} = (1 - \frac{T_2}{T_3})y_3 + \frac{T_2}{T_3}y_1 \quad (9)$$

where the algebraic variable $y_2$ is determined as:

$$y_2 = \begin{cases} P_C - P_{min}, & P_{min} > P_C - y_{2i} \\ P_C - P_{max}, & P_{max} < P_C - y_{2i} \\ y_{2i}, & P_{min} \leq P_C - y_{2i} \leq P_{max} \end{cases} \quad (10)$$

Note that, by the inclusion of the variable $u_c$, the governor speed reference in equation (6) is modified to $\omega_s - u_c$. As described in the next sections, the variable $u_c$ corresponds to the proposed coordination and is obtained by an ANN.

## III. CONTROL STRATEGY

The proposed coordination between SGs and WTs is described in this section. As discussed before, an slow response together with an small initial frequency deviation that gradually activates the governors lead to the typical frequency excursion observed in power systems dominated by thermal power plants. With an increasing penetration of wind power, WTs are urged to participate in frequency regulation [29]; still, due to the lack of energy storage, this participation can only be for a transient period. The issue with this transient response is that governors will see a transient improvement of the grid frequency and will not respond with the same intensity as if the grid were experiencing a larger frequency deviation. This results in either a further delayed response or a further decay in frequency when the control support from WTs is over. In this work we are proposing that, together with the grid frequency, an artificial coordinating loop (signal $u_c$) between SGs and WTs can prepare governors in advance for an enhanced frequency response without excessive overshooting. Intuitively, one might think that the coordinating signal $u_c$ can be derived based on the power reference of the WT's frequency controllers ($\Delta P$ in Fig. 2). In order to fully learn from this problem, we are proposing the use of an ANN that will provide an optimal signal. Here, we create a feed-forward back propagation network including one hidden layers and one output layer. The first layer has weights coming from the inputs and each subsequent layer has a weight coming from the previous layer. The hidden layer is added to process and identify the patterns among the data. The proposed ANN has the following general description:

1) *Target:* The main goal while training the network is to have a less severe frequency nadir after a disturbance. In steady-state, the grid frequency is unique; however, during transient state, generators will respond differently, setting distinctive bus frequencies throughout the grid. Moreover, due to electromechanical oscillations, some buses might exhibit more frequency changes and ultimately a larger frequency nadir than others. To tackle this issue, we use the system center of inertia to get a general frequency for the whole grid [30]. Minimizing frequency nadir implies getting rid of the overshooting during a frequency excursion. Note that the area $(S)$ of the frequency response when its value is only less than the steady state frequency is a good index for how well the frequency nadir is being reduced. As a result, the ANN goal is to minimize the computed area and make it as much as possible close to zero, which is the best achievable performance. The ANN target is set to zero as shown in Fig. 3, i.e., $S_{ref} = 0$. This reference value is not a practical target to track in reality, however, it can train the network in a fashion that can reduce the frequency nadir during the excursion.

2) *Input:* These correspond to the WTs active generated power, the SGs speed and computed error based on the above defined target, this is, $e = S - \text{target} = S$ as the target is zero.

3) *Output:* The actual output $y$ is the output of the last layer of the network. This output is defined to be the control signal $u_c$. The network is trained based on the expected output. Here, the expected output ($\hat{y}$) is considered as an improved signal based on the power reference of WT's frequency controllers, $\hat{y} = \alpha \Delta P$, where $\alpha$ is a constant parameter that represent our expected output. In this fashion, the neural network can be trained in a way to improve the primary frequency performance and reduce the frequency excursion.

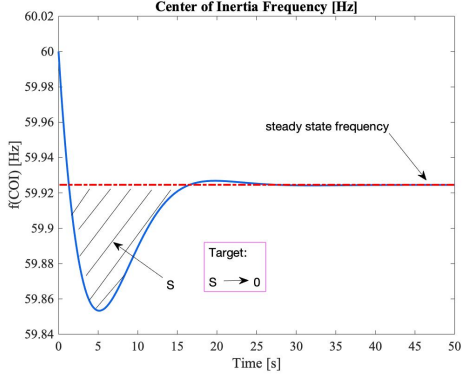

Fig. 3. Target for the neural network.

## IV. Application

The approach is applied to a 9-bus test system with a small wind farm connected to bus 8. The wind farm consists on 5 3.6 MVA DFIG-based WTs. Wind speed is assumed to be 11 m/s; under this condition, the wind farm power output is about 14 MW which corresponds to about 4.43% of the system total generated power (wind power penetration). System configuration and data is presented in reference [31]. All required parameters are presented in the Appendix. As disturbance, the load at bus 8 is increased 10% at time zero. The purpose of this application is to have a proof of concept, so the idea of creating a control coordination between WTs and SGs is validated. In further research, the coordination must be designed to include any potential disturbance in the system that causes power imbalances leading to frequency excursions.

### A. Training

In the provided network structure, a divide function is employed to separates data into training, validation, and testing subsets randomly. The ratio out of the 50,000 data points assigned to the training, validation and testing are 70%, 15% and 15%, respectively. This determined structure can evaluate ANN's performance across training, validation and testing data-sets repetitively. In this fashion, the ANN provides the lowest mean square errors (MSEs) with the given data-sets. The cost function ($C$) that generates the MSEs is assumed as:

$$C = \frac{1}{k} \sum_{i=1}^{k} \|\hat{y}_i - y_i\|^2 \tag{11}$$

where $k$ is the total number of training samples and $y_i$ is the vector of actual output from the network, known as $u_c$ in this framework. The MSE generates from the same cost function across training, validation, and testing data-sets that are strong indicators of ANN's overall performance [32]. Best validation performance for the trained networked is shown in Fig. 4 including the MSE's performance for the training, testing and validation data-sets. As shown, the ANN delivers impressive performance in generalizing the data as the differences between the values of MSE for training, testing and validation date-sets are insignificant.

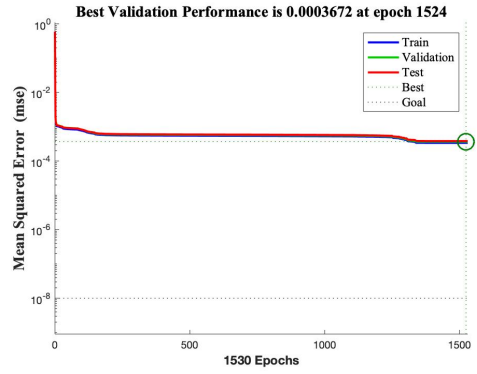

Fig. 4. Trained network performance.

Neural network training regression is illustrated in Fig.5. These regressions express the relationship between the output of the network and targets for the training, testing and validation sets. The solid lines in each of these figures show the best fit linear regression between output and targets that that rarely occurs in practice. The regression value $R$ is a statistical measure of how close the data is to the fitted regression line, that can be computed by a general formula described as [32]:

$$R = \sqrt{1 - \frac{\text{MSE of regression line}}{\text{MSE of the mean of the data}}} \tag{12}$$

As shown in all regression curves, all regression values are close to one, which means that the model illustrates all the variability of the output data around its mean.

### B. Controller design

The structure of the proposed feed-forward back-propagation network includes one input layer, one hidden layer and one output layer as it is shown in Fig. 6. The input layer receives data depending on the number of parameters defined in Section III. The weights of the network is computed by training the network using the back-propagation algorithm. The back-propagation algorithm is a supervised learning algorithm that performs a gradient descent search on the squared error energy surface to reach the minimum. [33]. Hence, the control coordinating signal $u_c$ can be obtained in order to improve the frequency excursion control by reducing frequency nadir as well as RoCoF after generator outages or load changes.

This effort can help improving governors response when the transient frequency support from WTs is over. Therefore, we can close the whole system by applying the vector of actual output from the neural network $u_c$, that is trained based on the collected data from the simulations. This method delivers a supplementary control loop for the grid that is able to enhance primary frequency regulation. This proposed novel idea creates a non-existent control coordination among renewable resources such as WTs and SGs which would be a transforming concept for actual power grids specially considering higher penetration of renewable energy.

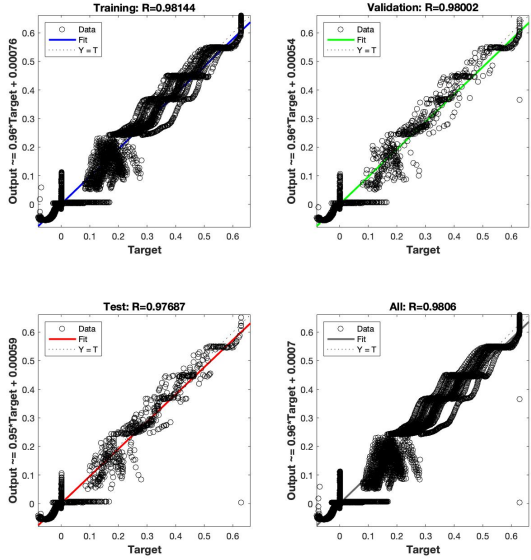

Fig. 5. Neural network training regression.

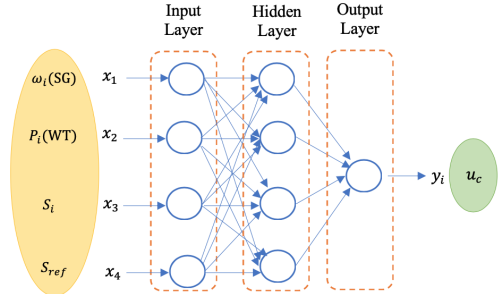

Fig. 6. The structure of feed-forward back-propagation neural network.

### C. Simulation results

The closed-loop system performance for primary frequency regulation in the 9-bus test system is presented in Fig. 7. The frequency of the three generators is shown for the cases of coordination and no coordination. Note that coordination allows reducing the frequency nadir in about 22%. Besides, the RoCoF has been improved as well from -0.0567 Hz/s to -0.04 Hz/s, which is about 29.5% of improvement. The illustrated results show that the proposed method could improve the frequency excursion control significantly by adding the coordinating mechanism between SGs and DFIG-based WTs. The provided control signal shown in Fig. 8 converges to zero after regulation is completed. Power output from the WTs based on its rotor speed in the closed-loop system is illustrated in Fig. 9. This figure shows less deviation in the generated power by applying the proposed control strategy, which is attractive considering the limited control capabilities of WTs.

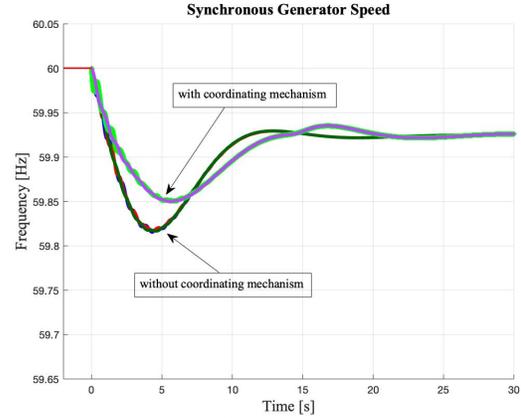

Fig. 7. Frequency of all synchronous generators in the grid

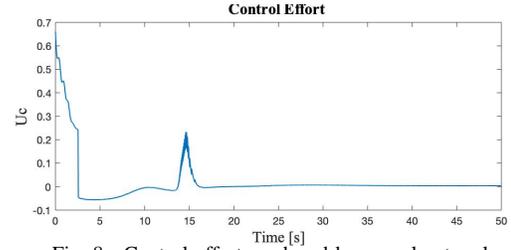

Fig. 8. Control effort produced by neural network

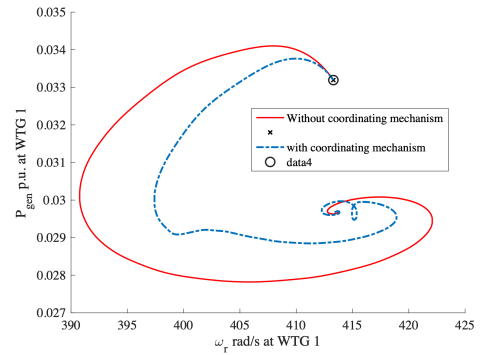

Fig. 9. Power from wind turbine vs rotor speed

## V. CONCLUSIONS

A novel coordinating mechanism between SGs and DFIG-based WTs for enhanced frequency regulation has been presented. The proposed coordination and concept behind it has a lot of potential for applications in actual power grids including different types of power plants, types of renewable energy resources, and control actuators. The coordination has been obtained through an ANN with a feed-forward back-propagation structure; the network consists in an input, hidden, and output layers. The proposed concept has been applied to a 9-bus test system. After a 10% load increase in bus 8, the results show that the proposed coordination allows reducing frequency nadir in about 22% and RoCoF in 29.5%. These comprehensive results validate our novel idea to improve frequency regulation through the coordination control between SGs and DFIG-based wind turbines using ANN. Further work is required to make this coordinating signal adaptive to any type of disturbance causing power imbalances. The development of coordinating signals can certainly strengthen power systems specially considering more aggressive penetration levels of renewable energy.

## APPENDIX

Wind turbine: $H_D = 1.23$, $X_m = 0.007$, $X_s = 3.37$, $X_r = 3.47$, $K_P = 0.398$, $K_I = 0.066$, $K_{P1} = 1$, $K_{P2} = 1$, $K_{P3} = 1$, $K_{P4} = 1$, $K_{I1} = 5$, $K_{I2} = 5$, $K_{I3} = 5$, $K_{I4} = 5$

Synchronous generator: $T'_d(SG_1) = 6$, $T'_d(SG_2) = 5.89$, $T'_d(SG_3) = 8.96$, $X_d(SG_1) = 0.8958$, $X_d(SG_2) = 1.3125$, $X_d(SG_3) = 0.1460$, $H(SG_1) = 6.4$, $H(SG_2) = 3.01$, $H(SG_3) = 23.64$.

IEESGO governor: $T_1 = 1$, $T_2 = 0.3$, $T_3 = 5$ $T_4 = 12$, $K_1 = 30$.

IEEE type-1 exciter: $K_A = 0$, $T_A = 20$, $K_E = -5$, $T_E = 1$, $K_F = 0.314$, $T_F = 0.063$.